\documentstyle[12pt,aasms]{article}
\def\etal{{\it et al.}}

\tightenlines

\begin{document}

\title{Anisotropy in the Microwave Sky at 90~GHz:\\  Results From Python~II}
\author{
J. E. Ruhl\altaffilmark{1,2},
M. Dragovan\altaffilmark{1,3},
S. R. Platt\altaffilmark{4},
J. Kovac\altaffilmark{5},
and
G. Novak\altaffilmark{6}
}

\altaffiltext{1}{Enrico Fermi Institute, Univ. of Chicago, Chicago, IL 60637}
\altaffiltext{2}{Current address: Dept. of Physics, Univ. of Calif. Santa
Barbara, Santa Barbara, CA 93106}
\altaffiltext{3}{Dept. of Astronomy and Astrophysics, Univ. of Chicago,
Chicago, IL 60637}
\altaffiltext{4}{Yerkes Observatory, Univ. of Chicago, Williams Bay, WI 53191}
\altaffiltext{5}{Dept. of Physics, Princeton Univ., Princeton, NJ 08544}
\altaffiltext{6}{Dept. of Physics and Astronomy, Northwestern Univ.,
		Evanston, IL 60208}

\begin{abstract}
  We report on additional observations of degree scale anisotropy at 90~GHz
from the Amundsen-Scott South Pole Station in Antarctica.
Observations during the first season with the Python instrument yielded a
statistically significant sky signal
with an amplitude of $\Delta T / T \sim 3.5 \times 10^{-5}$
for a Gaussian autocorrelation function model with a coherence
angle $\theta_c = 1^\circ$.  In this paper we
report the confirmation of that signal with data taken in the
second year, and on results from an interleaving set of fields.
Using the entire data set, we find
$\Delta T / T = \sqrt{C_0} = 2.8^{+1.1}_{-0.7} \times 10^{-5}$
for the Gaussian autocorrelation model mentioned above, and
$\Delta T / T = \sqrt{\ell_e(\ell_e+1)C_\ell/(2\pi)} =
2.1^{+0.7}_{-0.5} \times 10^{-5}$ for a band-power
estimate, where $\ell_e = 93$ is the effective center of our window
function.
The stated errors represent a 68\% confidence
interval in the likelihood added in quadrature with
a 20\% calibration uncertainty.
\end{abstract}


\section{Introduction}
Anisotropies in the Cosmic Microwave Background Radiation (CMBR)
contain a wealth of information about the conditions and processes
that led to the formation of large scale structures in the universe.
Whether the correct model of the evolution of the universe involves
baryons, exotic cold dark matter,
massive neutrinos, topological defects,
a standard ionization history or reionization at a modest
redshift, the content and history
of the universe affect the observed anisotropies of the
CMBR (see \cite{White} for a review).
Constrained by the measured level of anisotropy on large angular
scales~(\cite{DMR2,FIRS}), different models predict varying levels of
anisotropy on smaller scales.
Several observing teams~(\cite{MSAMI,ARGO,MAX,UCSB,SK}) including our own
have reported detections of anisotropy on angular scales near $1^\circ$.
Improvements in these measurements should
help discriminate between the current models of
large scale structure formation.

The Python instrument is designed to search for anisotropies
at an angular scale near $1^\circ$. The first observations (hereafter PyI)
with Python
were made between 1~and~15 January 1993 at the Amundsen-Scott South Pole
station~(\cite{PyI}, hereafter Paper~I).
Statistically significant signals were detected in observations
at high Galactic latitude
($|b| > 49^\circ$).

  We report here the results of a second set of
observations (hereafter PyII),
made between 12 and 23 December 1993 from the same site.
The bulk of the observing time in the second season was spent
on the PyI field (``Field~A"), in an effort to repeat that measurement.
The remainder was spent on a set of interleaving spots (``Field~B").
The observation of the Field~A spots serves as a further
check for the effects of interfering signals that would differ
from one year to the next,
including effects from atmospheric emission, cosmic ray hits in
the bolometers, and radiofrequency interference.
Additionally, the Sun was approximately $20^\circ$ further away from the
observed fields during the PyII season than it was for PyI.

\section{Instrument}
  The Python instrument~(Paper~I, \cite{Ruhlthesis}) consists of a 0.75~m
diameter off-axis parabolic
telescope that couples radiation into
a $2 \times 2$ array of bolometric detectors.
The beam response is well approximated by a Gaussian with
a full width at half maximum of $(0.75 \pm 0.05)^\circ$.
Radiation from the sky is first reflected off a vertical flat before
being focused into the cryostat by the primary.
Rotation of the flat about a vertical axis moves the detector response
horizontally across the sky.
The telescope is mounted on an
azimuth-elevation mount and is
surrounded by a large shield that protects the
instrument from being illuminated by the Sun or Earth.

Inside the cryostat there are four corrugated feed horns
at the focal plane
of the primary.  Radiation that enters a feed horn passes
through a set of single-mode waveguide filters
that define the passband at $\lambda = 3.3$~mm
before reaching the bolometric detectors.
The detectors use a layer of bismuth evaporated on a diamond wafer
as the absorbing element, and a chip of neutron transmutation doped
germanium as the thermistor element.  They are operated at
50~mK, cooled by a $^3$He guarded
adiabatic demagnetization refrigerator~(\cite{R_and_D}).
An additional bolometer (the ``dark channel"),
mounted on the cold stage but kept in a
sealed cavity, acts as a monitor for extraneous pickup.

New bolometers, constructed using the same methods as the
originals,
were installed for the PyII observing season.
However, two of the four new optical channels did not work well.
One channel was a factor of 3 less sensitive than typical
(6~mK$\cdot \sqrt{\mbox{s}}$ instead of 2~mK$\cdot \sqrt{\mbox{s}}$);
the other was unusable due to an electrical problem in the dewar.
Fortunately, one good channel was on the upper row, and the other was on
the lower, making possible the confirmation of
the bulk of the PyI data.

The two most significant changes to the telescope between PyI and PyII
were the improved balancing of the chopping mechanism for the
external switching flat, and the installation of a microwave-absorbing
guard ring around the primary.

\section{Observations}
The Sun, Moon, and two sources in the Carinae nebula were used
as absolute pointing references each
season;  additionally, Venus was used during the PyI season.
We estimate our absolute pointing accuracy to be $\pm 0.1^\circ$
each year, or $\pm 0.15^\circ$ for the relative pointing
accuracy of the PyI and PyII datasets.

The same observing strategy was used for PyI and PyII, and is
described in Paper~I.  The combination of a fast 3-beam chop
(2.5~Hz full cycle) and slower beamswitching (10 to 30 seconds per position)
yields a 4-beam response on the sky,
\begin{equation}
\Delta T_{j}^k =  -\frac{1}{4} T_j + \frac{3}{4} T_{j+1}
                    -\frac{3}{4} T_{j+2} + \frac{1}{4} T_{j+3},
\end{equation}
where the $T_j$'s are the antenna temperatures of patches on
the sky separated by $2.75^\circ$ along a horizontal line.
A scan consists of measuring $\Delta T_j^k$ three times
successively~($k=1,2,3$) at each of 7 positions $j$
on the sky.  The left and right hand channels in a given row of the array
measure many of the same 4-beam patterns;  the spots measured by
the left hand side are given by $j=1,...,7$, while those made by the
right correspond to $j=2,...,8$.

The time-ordering of the right and left-hand
3-beam measurements (``stares'') are
reversed from one value of $j$ to the next.  For the first 4-beam pattern
($j=1$ for the left channels, $j=2$ for the right),
the left stare is measured first.  For the second
4-beam pattern the right stare is measured first,
and so on.
This causes a drifting 3-beam offset to appear as an
oscillating 4-beam signal on the sky.
However, the measured linear drifts in the 3-beam offset
within each scan are both small and symmetric about zero, making
this effect unimportant.
The average 3-beam offsets are less than 1~mK for
all the PyI and PyII channels,  and the average slopes are
within $2\sigma$ of zero, with an error of $\sigma \sim 1 \mu$K/stare.

\section{Calibration}
  The PyI observations were calibrated by a combination
of elevation scans and the placement of known-temperature
blackbody and low emissivity foam loads
in the optical path, as described in Paper~I and \cite{R_capri}.
The PyII observations were calibrated using the foam loads;
the loads were in turn calibrated by placing a
large liquid-nitrogen-cooled load beneath the
dewar and alternately switching the foam and two known-temperature loads into
the optical path.  This calibration
of the foam loads was done {\it in situ} on the telescope during
the PyII season,
rather than in a 300~K room, reducing possible systematic effects
in the measurement.  The DC gain measured
using the foam loads was converted to an AC gain
by comparing the signals seen while switching on and off a foam
load slowly (0.1~Hz) with those seen while switching at the frequency used
for the observations (5~Hz).

The relative calibrations of the various bolometer
channels (four channels for PyI, three for PyII)
were checked using two sources in
the Carinae nebula;  the results from all seven channels
lie within 15\% of the average.
The two sources in the Carinae nebula are at
($\alpha$,$\delta$) = $(10^{\mbox{\scriptsize h}}44^{\mbox{\scriptsize m}},
-59.64^\circ)$ and
$(10^{\mbox{\scriptsize h}}33^{\mbox{\scriptsize m}},-57.95^\circ)$
in J1994 coordinates.
The mean signals from them are
($9.0 \pm 0.3$) and ($4.2 \pm 0.1$)~mK in our beam, respectively,
with the errors representing only the statistical uncertainty.
The first source was also used as a pointing reference.

The statistical error on the determination of the average gain
is 5\%.  Our estimates of possible systematic
effects are at the level of $<20$\%.  We therefore adopt a
gain uncertainty of 20\%.

\section{Analysis}
  The data analysis presented here differs slightly from
that used for the previously
reported PyI results; using the new method does not significantly change
those results.  We describe here the new method, which is used to arrive at
all the results in this paper.

The output of each detector is sampled at 100~Hz
and demodulated in software, synchronously with the
motion of the chopping flat.
Two lockin demodulations are used;  one (the ``optical phase") is
maximally sensitive to the 3-beam signal from the sky, while the second
(the ``quadrature phase") is shifted by $90^\circ$ from the
optical phase and should have no response to a stationary sky signal.
We confine our discussion to the optical phase
data, giving results from the quadrature phase where relevant.

Within each of the 42 stares in a scan,
the mean and variance
$\sigma_{S_j}^2$ of the lockin values are calculated.
The means are then combined in a pairwise fashion into 4-beam values
$\Delta T_j^k$ as described above, and an average variance
$\sigma_{S}^2$ is calculated from the 42 values of $\sigma_{S_j}^2$.
Thus $\sigma_{S}^2$ is a measure of the noise which contains drifts
only within 10 to 30 second long stares.

As previously described, each channel measures a 4-beam value
$\Delta T_j^k$ three times in succession $(k = 1,2,3)$
for seven sky positions $j$.  From these we find
a mean for each channel at each position $j$,
and an average error on those means, $\sigma_m$.
Drifts up to the
time separating observations of successive 4-beam positions
(roughly 3 minutes), are included in $\sigma_m$.
The variance from a celestial signal does not contribute to
$\sigma_{m}$, and $\sigma_{m}$ can therefore be used as an unbiased
statistic for cutting scans.

The noise estimates described above are calculated for each channel
in every complete scan.
We use $\sigma_{S}$ and $\sigma_{m}$, normalized to
a ``stare" duration of 30 seconds,
on a channel by channel basis to remove scans which have
been contaminated by excessive noise.
The first cut removes scans with $\sigma_{S}$
greater than 1.5 times the peak of
its distribution.  This procedure removes 20 to 25\% of the
scans for each channel in PyI, and 5 to 10\% of those in PyII.
The second cut removes scans with high values of $\sigma_{m}$;
the value at which scans are cut is placed so as to minimize the
average final errorbar in the binned data.
This step removes 19 to 27\% of the scans for PyI, and
4 to 5\% of those in PyII.  The cuts remove more PyI data than PyII
because of the poorer weather during that season.
In all, approximately 50 hours (depending on the channel) of PyI data
passed the cuts.  For PyII, roughly 24 hours of data on the overlapping
field (Field~A), and 9 hours on the interleaving field (Field~B)
remained after the cuts.

The average value of $\sigma_{S}$ from all scans that pass the cuts
is a good indicator of chopped detector noise.  For PyI, the noise
in the four channels is 2.5, 2.9, 2.1 and 1.9 mK~$\cdot \sqrt{\mbox{s}}$;
For the three operational channels of PyII, it is
2.1, 1.3 and 5.7 mK~$\cdot \sqrt{\mbox{s}}$.
These values are in good agreement with those found
from the quadrature values of both $\sigma_{S}$ and $\sigma_{m}$.
The optical phase value of $\sigma_{m}$ is higher (for the
sensitive channels), due to atmospheric contamination.
For PyI it is 4.1, 4.2, 4.0 and 3.9 mK~$\cdot \sqrt{\mbox{s}}$;
for PyII these values are 2.9, 2.7 and 5.6 mK~$\cdot \sqrt{\mbox{s}}$.

For sky positions observed by more than one channel,
the 4-beam averages from neighboring
channels are combined into a single value within a scan
by forming a weighted average of the
left and right-hand channels.  The average value of $1/\sigma_S^2$ over
all uncut scans is used as the weight.
If one of the channels is cut by the previously described
procedures, its neighbor is cut as well.

After this treatment,
a statistical mean and error are calculated from the uncut scans
at each of the sky positions $j$.
The errors are within 10\% of those expected given the distribution
of $\sigma_m$, indicating there is little if any additional
atmospheric contamination.
These means and errors, after multiplying by 1.07 to
correct for the estimated
atmospheric absorbtion in our band, are the final values and errors
for the 4-beam temperature differences on the sky.

\section{Results}
The results of the analysis of the PyI data
and of the overlapping portion (Field~A) of the PyII data set
appear in Figure~\ref{GE_Ts}.
The agreement is good; taking the difference of the two data sets
[ $(\mbox{PyI} - \mbox{PyII})/2$ ]
gives a result that agrees well with zero signal
(reduced $\chi^2 \equiv \chi^2_\nu = 12./15$).
The weighted mean of the two data sets is shown
as the filled circles in Figure~\ref{GEGC_Ts};
here the agreement with zero signal is very poor ($\chi^2_\nu = 191./16$).
The contrast between the $\chi^2$'s for the summed and differenced
data sets indicates that the signal in the two data sets is the same.

A set of patches (Field~B) that interleaves those from Field~A was also
observed during the PyII season.  The locations of the
Field~B beams are found by moving one half of a chopper throw
in negative right ascension ($1.38^\circ$ on the sky) from the beams
that make up Field~A.
Analyzing the Field~B data
leads to the values shown as open squares in Figure~\ref{GEGC_Ts}.
Less time was spent observing this field,
and the errors are larger than on Field~A.
The agreement with zero sky signal is
good for Field~B ($\chi^2_\nu = 14./15$).

The results shown in Figure~\ref{GEGC_Ts} were measured in a single
frequency band, so we cannot spectrally discriminate between CMBR
anisotropies and the various possible foregrounds.  In Paper~I we
discussed the expected levels of galactic foreground contamination,
which are smaller than the signals seen.  Proceeding under the
assumption that these signals represent fluctuations in
the CMBR, we multiply the values plotted in
Figure~\ref{GEGC_Ts} by 1.24 to convert them from
Rayleigh-Jeans temperature units to temperature differences
in a 2.73~K blackbody, and use
the integrated likelihood function to form
confidence interval estimates for the sky signal in two models
with Gaussian CMBR fluctuations.  For both analyses we use
the full correlation matrix including off-diagonal elements
describing correlations present because of the theoretical model and
those induced by the observing strategy.

  The first model consists of a sky with a Gaussian autocorrelation
function~(GACF) with a coherence angle of $\theta_c = 1^\circ$.
We set limits on $\sqrt{C_0}$, the rms amplitude of the fluctuations in
$\Delta T/T$.  The uncertainty in beamwidth leads to less than
a 2\% error in $\sqrt{C_0}$.
The GACF limits can be converted to an estimate
of the flat band power sampled by our window function.  Following the recipe
of \cite{Bond1}, we find for our window function,
$\Delta T_{\mbox{\scriptsize \it Band Power}} = 0.73\Delta T_{GACF}$.
The effective center of our window function
for a flat band power spectrum lies at $\ell_{e} = 93$.

We also use a model with an uncorellated sky,
$C(\theta) = \delta(1-\cos\theta)$,
to find the rms of our
data set, from which we derive another band power estimate.
Correlations due to beam overlap are less than 1\%, and we ignore them.
This analysis gives results that are within a few percent of those
found by converting
the GACF limits, indicating that the flat band power is insensitive
to the form of the corellation function used in the likelihood
calculation.

The results for the various combinations of data sets
and analyses are given in Table~\ref{table_res}.
The confidence intervals quoted in the table do not include the
20\% calibration error.  The first four
entries in Table~\ref{table_res} show that the
signal seen in the first season appears again in the second.
The last entry in Table~\ref{table_res} gives our final result
using the data from both seasons and both fields.

\section{Conclusions}
After detecting a signal in our first year of observations, we
re-observed the same portion of the sky and detected the same signal.
This provides further evidence that the signal is
celestial rather than systematic in nature.

Additionally, an interleaving set of spots was
observed during the second season.
Likelihood analyses on the entire data set are used to
derive two estimates of the sky signal.  First, we find
$79^{+28}_{-19} \mu$K for the total sky rms for a
Gaussian autocorrelation model with a coherence angle of $1^\circ$.
Second, we find a band power of $57^{+20}_{-14} \mu$K,
using an autocorrelation fuction given by
$C(\theta) = \delta(1 - \cos\theta)$.
The stated errors are 1~$\sigma$ limits that include the 20\% calibration
uncertainty added in quadrature with the likelihood-derived
errors.

We thank the Antarctic Support Associates staff at the South Pole
for making a successful season possible, and
Ted Griffith, Bob Pernic, and Bill Vinje
for valuable assistance there.  This research was supported by the
James S. McDonnell Foundation,
PYI grant NSF AST 90-57089, and
the National Science Foundation
under a cooperative agreement with the Center for Astrophysical Research in
Antarctica (CARA), grant NSF OPP 89-20223. CARA is an NSF
Science and Technology Center.  JR was supported by the McCormick
Fellowship at the University of Chicago.

\newpage
\begin{table}[h]
\centering
\begin{tabular}{c c c c}   \hline \hline
Dataset& Field         & Band Power         & GACF             \\
&	& $T_0 \sqrt{\ell_e(\ell_e+1)C_\ell/(2\pi)}$ & $T_0 \sqrt{C_0}$ \\
&		& $\ell_{e} = 93$     & $\theta_c = 1^\circ$ \\ \hline
PyI          & A& $66^{+27}_{-11}$   & $90^{+36}_{-15}$ \\
PyII         & A& $68^{+32}_{-11}$   & $91^{+42}_{-15}$ \\
(PyI\&PyII)  & A& $69^{+27}_{-10}$   & $93^{+36}_{-13}$ \\
(PyI-PyII)/2.& A& $(3,19) $              & $(5,26)$     \\
PyII         & B& $(16,59)$              & $(23,81)$  \\
(PyI\&PyII)&A\&B& $57^{+16}_{-8}$   & $79^{+23}_{-10}$ \\ \hline
\end{tabular}
\centering{
\parbox[htb]{5in}{
\caption{ Results of likelihood analyses.
\label{table_res}}
Values are in units of $\mu$K.
Detections are quoted at the maximum in the likelihood,
with 16\% of the integrated likelihood above the upper errorbar,
and 16\% of the integrated likelihood below the lower one.
For datasets with no significant detection, the value
which maximizes the likelihood is given, along with a
95\% upper limit from the integrated likelihood.
The 20\% calibration uncertainty is not included in these errors.
The signals from the dark channel and from the quadrature
phase are consistent with zero.
}
}
\end{table}

\newpage
\begin{figure}[b]
\vspace{1in}
\caption{
Results from PyI and PyII for Field A.  The PyI results are
plotted as the solid circles, and the PyII results are
shown as the open squares.  They are offset horizontally
for clarity.  The 4-beam response pattern is shown in the top
panel, aligned with the point plotted at $\alpha = -12^\circ$.
The temperature axis is plotted in Rayleigh-Jeans units;
to convert to thermodynamic temperature
differences for a 2.73~K blackbody, multiply by 1.24.
\label{GE_Ts}
}
\end{figure}

\begin{figure}[b]
\vspace{1in}
\caption{
Final results for both fields.  The weighted mean of
PyI and PyII for Field~A are shown as the solid circles.
The open squares show the results of the observations of
Field~B done during the PyII season.
The temperature axis is plotted in Rayleigh-Jeans units;
to convert to thermodynamic temperature
differences for a 2.73~K blackbody, multiply by 1.24.
\label{GEGC_Ts}}
\end{figure}

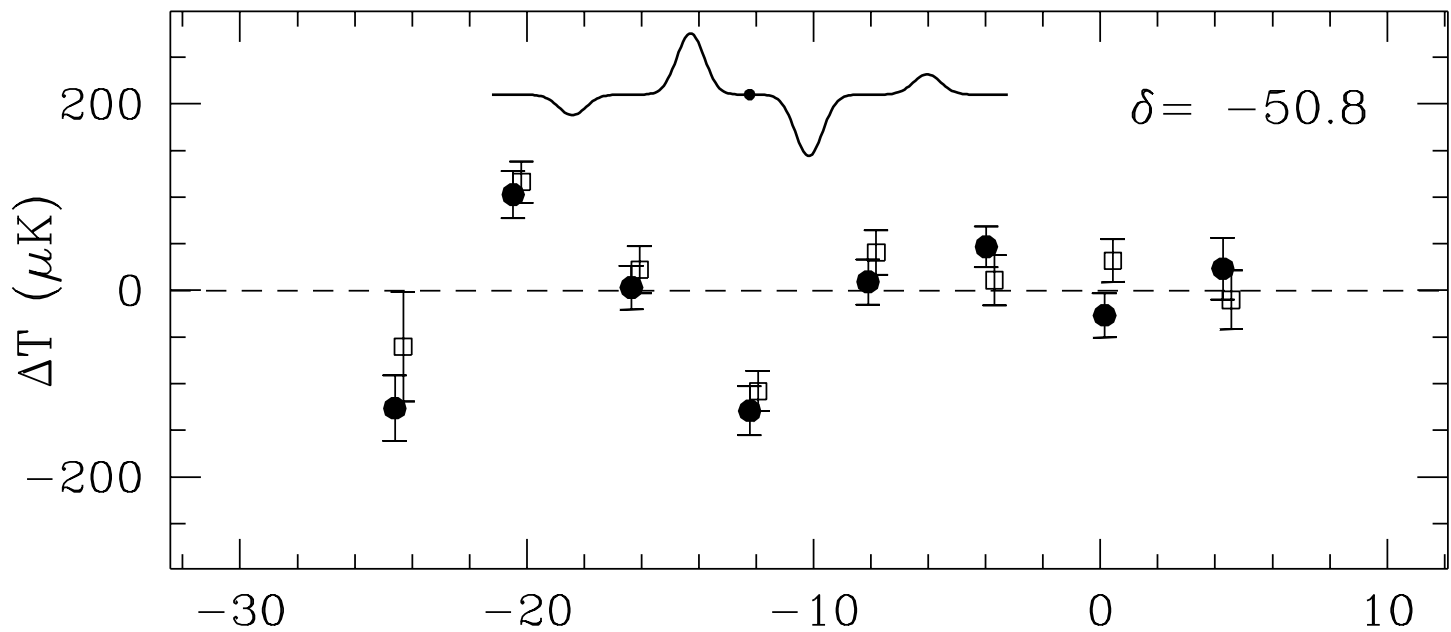
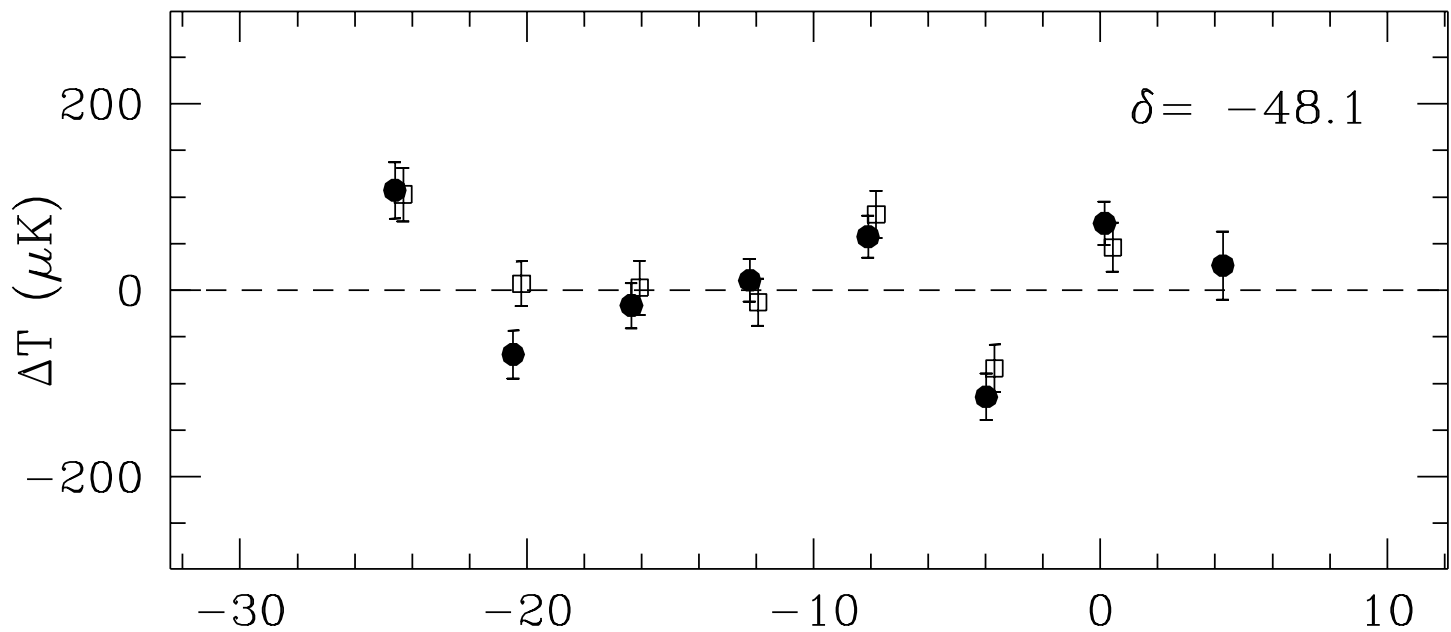
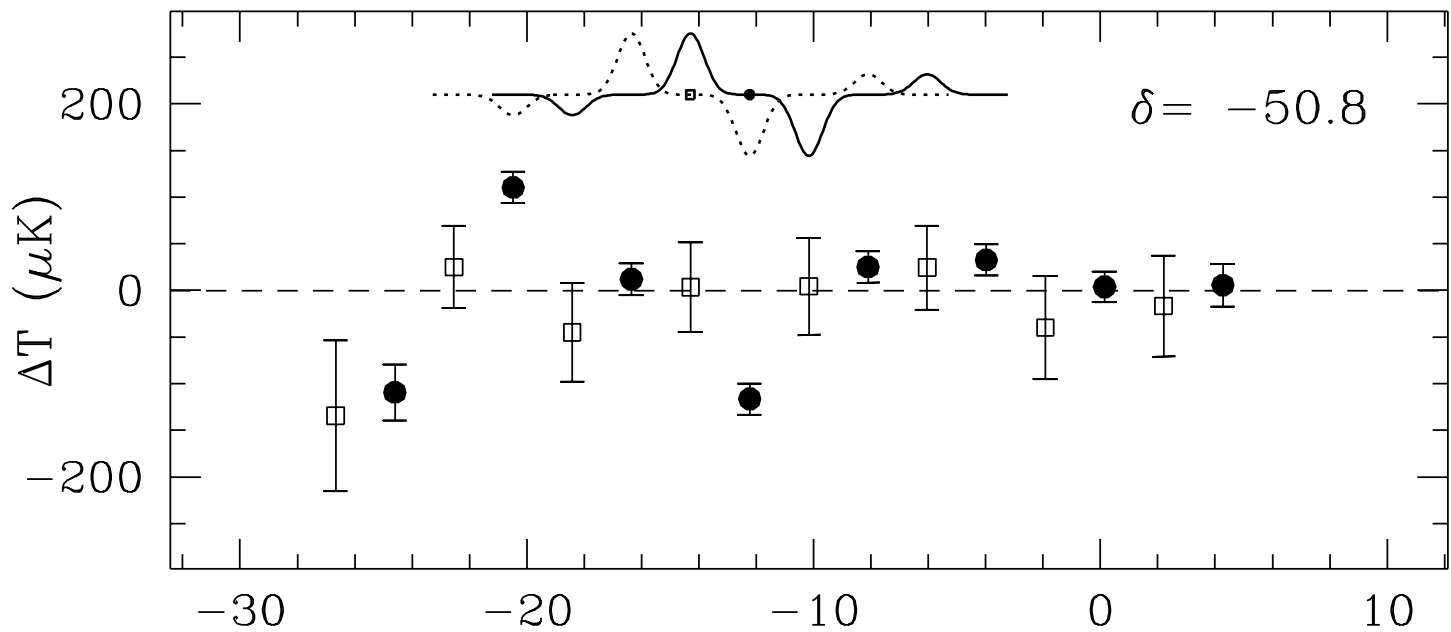
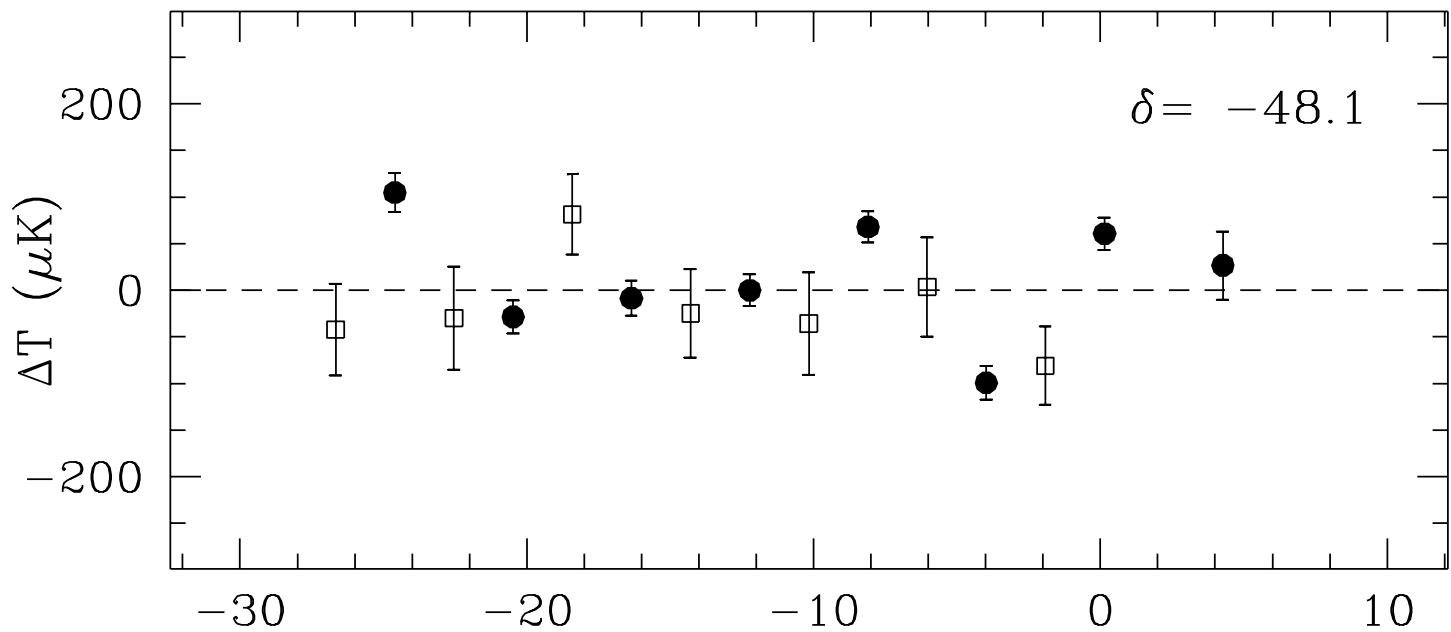
\end{document}